\documentclass{ifacconf}

\usepackage{graphicx}      
\usepackage{natbib}        

\usepackage{amsmath}
\usepackage{amssymb}
\usepackage{amsfonts}

\usepackage{algorithm}
\usepackage{algpseudocode}

\usepackage{multirow}
\usepackage{booktabs}

\usepackage{url} 
\usepackage{enumitem} 
\usepackage{glossaries}

\newacronym{mpc}{MPC}{model predictive control}
\newacronym{simd}{SIMD}{single instruction, multiple data}
\newacronym{blas}{BLAS}{basic linear algebra subprograms}
\newacronym{bblas}{BBLAS}{batched BLAS}
\newacronym{lapack}{LAPACK}{linear algebra package}
\newacronym{cr}{CR}{cyclic reduction}
\newacronym{kkt}{KKT}{Karush-Kuhn-Tucker}
\newacronym{ipm}{IPM}{interior-point method}
\newacronym{spd}{SPD}{symmetric positive definite}
\newacronym{rts}{RTS}{Rauch--Tung--Striebel}
\newacronym{rhs}{RHS}{right-hand side}
\newacronym{fp64}{FP64}{double precision}
\newacronym{fp32}{FP32}{single precision}
\newacronym{map}{MAP}{maximum a posteriori}

\newcommand{\tbdgpu}{\textbf{BlockDSS}}

\begin{document}
\begin{frontmatter}

\title{Harnessing Batched BLAS/LAPACK Kernels on GPUs for Parallel Solutions of Block Tridiagonal Systems\thanksref{footnoteinfo}} 

\thanks[footnoteinfo]{We gratefully acknowledge generous funding provided by Charles Cahn. We also thank Ryan Senne for valuable discussions on the neuroscience applications of this work.}

\author[First]{David Jin} 
\author[Second]{Alexis Montoison} 
\author[Third]{Sungho Shin}

\address[First]{Center for Computational Science and Engineering, Massachusetts Institute of Technology, 
   Cambridge, MA 02139 USA (e-mail: jindavid@mit.edu)}
\address[Second]{Math and Computer Science Division, Argonne National Laboratory, 
   Lemont, IL 60439 USA (e-mail: amontoison@anl.gov)}
\address[Third]{Department of Chemical Engineering, Massachusetts Institute of Technology, Cambridge, MA 02139 USA  (e-mail: sushin@mit.edu)}

\begin{abstract}


Block-tridiagonal systems are prevalent in state estimation and optimal control, and solving these systems is often the computational bottleneck. 
Improving the underlying solvers therefore has a direct impact on the real-time performance of estimators and controllers. 
We present a GPU-based implementation for the factorization and solution of block-tridiagonal \gls*{spd} linear systems. 
Our method employs a recursive Schur-complement reduction, transforming the original system into a hierarchy of smaller, independent systems that can be solved in parallel using batched BLAS/LAPACK routines. 
Performance benchmarks with our cross-platform (NVIDIA and AMD) implementation, \tbdgpu{}, show substantial speed-ups over state-of-the-art CPU direct solvers, including CHOLMOD and HSL MA57, while remaining competitive with NVIDIA cuDSS. 
At the same time, the current implementation still invokes batched routines sequentially at each recursion level, and high efficiency requires block sizes large enough to amortize kernel launch overhead.

\end{abstract}


\end{frontmatter}

\section{Introduction}\label{sec:intro}
We consider block-tridiagonal linear systems of the following form:
\begin{equation}\label{eq:btd-structure}
\underbrace{\begin{bmatrix}
A_{1,1}     & A_{2,1}^\top     &                 &         \\
A_{2,1}     & \ddots     &   \ddots       &         \\
            & \ddots  & \ddots  & A_{N,N-1}^\top \\
            &         & A_{N,N-1} & A_{N,N}
\end{bmatrix}}_{A}
~
\underbrace{\begin{bmatrix}
X_1 \\ X_2 \\[-0.8ex] \vdots \\ X_{N-1} \\ X_N
\end{bmatrix}}_{X}
=
\underbrace{
\begin{bmatrix}
B_1 \\ B_2 \\[-0.8ex] \vdots \\ B_{N-1} \\ B_N
\end{bmatrix}}_{B},
\end{equation}
where $A \in \mathbb{R}^{Nn \times Nn}$ denotes a symmetric positive definite (SPD) block-tridiagonal matrix composed of $N$ diagonal blocks $A_{i,i} \in \mathbb{R}^{n \times n}$ and $N-1$ off-diagonal blocks $A_{i, i+1} \in \mathbb{R}^{n \times n}$. The solution $X \in \mathbb{R}^{Nn \times d}$ and the \gls*{rhs} $B \in \mathbb{R}^{Nn \times d}$ are partitioned conformally into $N$ submatrices, each containing $n$ rows.

The \gls*{spd} block tridiagonal systems commonly arise in estimation and control contexts:
\begin{itemize}[leftmargin=*,itemsep=0pt,parsep=0pt,partopsep=0pt]
\item \textbf{Kalman Smoothing:}
  Kalman smoothing estimates the state of a dynamical system based on noisy measurements over a given time horizon.  
  Under the linear-Gaussian setting, the \gls*{map} estimator penalizes both the prediction and measurement errors quadratically, leading to a sparse normal equation.  
  The \gls*{rts} smoother \citep{Rauch1965} yields a block-tridiagonal \gls*{spd} Hessian matrix \citep{ARAVKIN2021821}, which can be efficiently solved using specialized block-tridiagonal solvers.  
\item \textbf{\Gls*{mpc}:}
  The \gls*{mpc} method computes control actions by repeatedly solving finite-horizon optimal control problems in real time.    
  The associated \gls*{kkt} systems inherit a block-tridiagonal structure from the temporal coupling of the dynamics.  
  Although these systems are often indefinite, condensation strategies can transform them into \gls*{spd} block-tridiagonal systems  \citep{pacaudGPUacceleratedDynamicNonlinear2024}.  
  GPU solvers are therefore attractive for achieving the performance required for real-time control \citep{pacaudGPUacceleratedDynamicNonlinear2024,du2025gatogpuacceleratedbatchedtrajectory}.  
\end{itemize}

This paper presents a scalable algorithm for solving the system in (\ref{eq:btd-structure}) by exploiting GPU parallelism.
A straightforward approach would be to apply classical tridiagonal algorithms, such as the Thomas algorithm, at the block level \citep{Aravkin2014}.
However, these schemes are inherently sequential, with each block depending on its predecessor, and a naive GPU implementation that iterates over blocks issues many small \gls*{blas} and \gls*{lapack} calls.
Even when the blocks are moderately sized, the computational cost per call may be too small to amortize kernel launch overhead, leading to underutilization and poor performance.

\begin{figure}[t]
  \centering
  \includegraphics[width=\columnwidth]{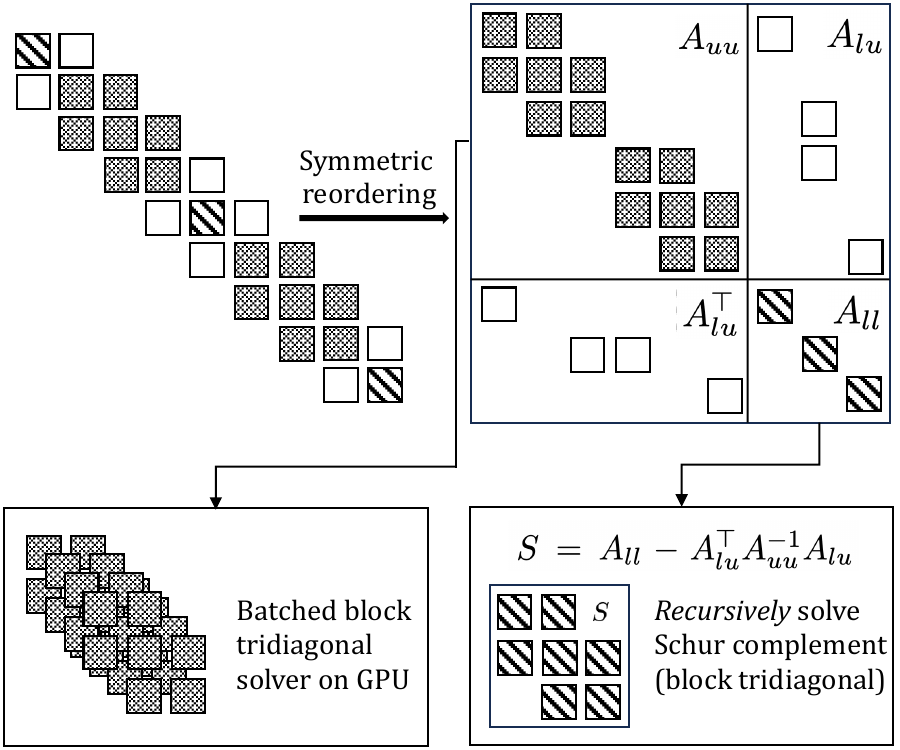}
  \vspace{-5mm}
  \caption{An overview of recursive Schur complement strategy for solving block-tridiagonal systems.}
  \label{fig:teaser}

\end{figure}

To address this limitation, we adopt a recursive Schur-complement strategy, also known as \emph{substructuring} in the domain decomposition literature \citep{Smith1996,TEREKHOV2013245} (see Fig. \ref{fig:teaser}).
Related formulations have been studied for block-tridiagonal systems \citep{belov2017}, but without explicitly targeting batched \gls*{blas} and \gls*{lapack} kernels, which are central to our approach.
A symmetric permutation reorganizes the matrix into a set of independent interior blocks and a set of separator variables.
The interior segments can then be eliminated in a single, massively parallel stage using batched kernels. Then, the resulting smaller, denser, and still block-tridiagonal Schur complement system is solved by recursively applying the same procedure.
The recursive procedure transforms a sparse, latency-bound problem into a hierarchy of dense, compute-bound kernels that are well suited to GPU execution.

The algorithm is expressed entirely in terms of batched \gls*{blas}/\gls*{lapack} routines.
Batched variants execute the same linear-algebra operation on many independent matrices, reducing kernel launch overhead and improving GPU utilization \citep{rennichAcceleratingSparseCholesky2014}.
In our case, the recursive Schur-complement strategy is implemented using only three standard batched routines: Cholesky factorization (\texttt{potrf}), triangular solve (\texttt{trsm}), and matrix--matrix multiplication (\texttt{gemm}), all widely available in vendor and open-source GPU libraries.

\subsubsection{Contribution} 
This paper introduces a recursive Schur-complement strategy tailored to \gls*{spd} block-tridiagonal systems and formulated explicitly in terms of batched \gls*{blas}/\gls*{lapack} operations on GPUs. While related nested Schur-complement ideas have appeared in dynamic programming and other contexts \citep{wrightPartitionedDynamicProgramming1991}, to the best of our knowledge they have not been developed for \gls*{spd} block-tridiagonal systems within a batched dense linear-algebra framework, nor coupled to structure-aware reordering for high GPU utilization. We further provide an open-source GPU implementation, \tbdgpu{}\footnote{https://github.com/MadNLP/BlockDSS.jl.git}, supporting both CUDA and ROCm backends and both single and double precision, which enables a systematic empirical comparison against state-of-the-art GPU and CPU sparse direct solvers and demonstrates the practical benefits of structure-exploiting batched kernels on modern hardware.

\section{Related Work}

\subsubsection{Classical and parallel tridiagonal algorithms}
The Thomas algorithm is the standard serial method for tridiagonal systems \citep{Thomas1949} (i.e., \(n = 1\) in (\ref{eq:btd-structure})) with time complexity \(O(N)\).  
To expose more parallelism, schemes such as cyclic reduction \citep{hockney1965cr} and block cyclic reduction \citep{sweet1977} were developed.  
These methods offer substantial concurrency but often at the cost of additional arithmetic work or reduced numerical robustness; see \cite{GanderGolub1997CyclicReduction} for a comprehensive survey.  
Block-tridiagonal extensions of these algorithms are important in applications such as Kalman smoothing and optimal control.

\subsubsection{GPU math libraries and batched kernels}
Modern GPU solvers heavily rely on vendor-provided dense linear-algebra libraries.  
NVIDIA provides cuBLAS for dense operations, cuSOLVER for factorizations, and cuDSS for sparse direct methods \citep{NVIDIACUBLAS2025,NVIDIACUSOLVER2025,NVIDIACUSPARSE2025,NVIDIACUDSS2025}; AMD offers analogous functionality through its ROCm ecosystem \citep{ROCm2025}.  
These routines are highly tuned to the underlying hardware and enable high-performance linear algebra on large matrices. Batched \gls*{blas}/\gls*{lapack} kernels execute the same operation over many independent small matrices, amortizing kernel-launch overhead and sustaining high GPU utilization; the BBLAS interface and its implementations in MAGMA and vendor libraries have made this style of computation standard \citep{dongarra2017bblas, magma2024}.  Our algorithm is formulated directly in this batched setting, using these primitives to perform block elimination at high throughput.


\subsubsection{Block-tridiagonal solvers on GPU}
Several libraries target tridiagonal or block-tridiagonal systems on GPUs.  
TridiGPU provides bandwidth-optimized implementations of Thomas and PCR variants \citep{klein2023tridigpu}, while PaScaL\_TDMA implements a hybrid Thomas/PCR approach for CUDA \citep{kkpc2020}.  
The PfSolve framework generalizes cyclic reduction strategies across multiple architectures, including GPUs \citep{tolmachev2025pfsolve}.  
Block-tridiagonal routines are also available in the ROCm library stack \citep{ROCm2025}, although, as of ROCm v6.4.1, these routines do not yet handle all problem sizes robustly. These solvers achieve excellent performance for tridiagonal or narrow-banded systems, but they do not exploit larger block sizes, hierarchical Schur-complement factorizations, or tensor-core–optimized dense kernels required for large collections of block-tridiagonal \gls*{spd} systems.  

\section{Parallel Recursive Algorithm}\label{sec:parallel-algo}

\subsection{Partitioning and Reordering}
The core of the recursive algorithm lies in partitioning the original block-tridiagonal system into smaller, independent subsystems.
The key to the recursive method is to partition the $N$ block variables into two sets: a set of $P$ \emph{separators} and a set of $N-P$ \emph{interiors}. A symmetric permutation is applied to the entire system, which rearranges the matrix, the solution matrix $X$ into a $2 \times 2$ block structure.

We illustrate this procedure with a system with $N=9$ and $P=3$, where separators are at indices $\{1, 5, 9\}$.
The reordered matrix is given as follows:
\begin{align}
\label{eq:permuted-matrix}
  \begin{aligned}
     \widetilde A &=
         \left(
         \begin{array}{c|c}
            A_{uu}\phantom{^\top}\!\!\!\! &  A_{lu} \\
           \hline
            A_{lu}^{\top\phantom{^\top}}\!\! &  A_{ll}
         \end{array}
         \right) \\
    &= \left(
         \begin{array}{cccccc|ccc}
           A_{2, 2} & A_{3, 2}^\top & 0 & 0 & 0 & 0 & A_{2,1} & 0 & 0 \\
           A_{3, 2} & A_{3, 3} & A_{4,3}^\top & 0 & 0 & 0 & 0 & 0 & 0 \\
           0 & A_{4,3} & A_{4,4} & 0 & 0 & 0 & 0 & A_{5,4}^\top & 0 \\
           0 & 0 & 0 & A_{6,6} & A_{7,6}^\top & 0 & 0 & A_{6,5} & 0 \\
           0 & 0 & 0 & A_{7,6} & A_{7,7} & A_{8,7}^\top & 0 & 0 & 0 \\
           0 & 0 & 0 & 0 & A_{8,7} & A_{8,8} & 0 & 0 & A_{9,8}^\top \\
           \hline
           A_{2,1}^\top & 0 & 0 & 0 & 0 & 0 & A_{1,1} & 0 & 0 \\
           0 & 0 & A_{5,4} & A_{6,5}^\top & 0 & 0 & 0 & A_{5,5} & 0 \\
           0 & 0 & 0 & 0 & A_{9,8} & 0 & 0 & 0 & A_{9,9}
         \end{array}
         \right).
  \end{aligned}
\end{align}
The same permutation partitions the solution matrix $X$ and RHS matrix $B$ into components corresponding to the interiors ($ X_{u},  B_{u}$) and separators ($ X_{l},  B_{l}$):
\begin{equation}
\label{eq:permuted-vectors}
 \widetilde X = 
\begin{pmatrix}  X_{u} \\  X_{l} \end{pmatrix}
=
\left(
\begin{array}{c}
  X_2 \\ X_3 \\ X_5 \\ X_6 \\ \hline X_1 \\ X_4 \\ X_7
\end{array}
\right)
, \quad
 \widetilde B = 
\begin{pmatrix}  B_{u} \\  B_{l} \end{pmatrix}
=
\left(
\begin{array}{c}
  B_2 \\ B_3 \\ B_5 \\ B_6 \\ \hline B_1 \\ B_4 \\ B_7
\end{array}
\right)
\end{equation}
The reordered system $ \widetilde A \widetilde X =  \widetilde B$ can now be written as a coupled $2 \times 2$ block system of equations:
\begin{subequations}
  \begin{align}
     A_{uu}  X_{u} +  A_{lu}  X_{l} &=  B_{u} \label{eq:reordered-eq1} \\
     A_{lu}^\top  X_{u} +  A_{ll}  X_{l} &=  B_{l} \label{eq:reordered-eq2}
  \end{align}
\end{subequations}
The blocks of this reordered system can be interpreted as follows:
\begin{itemize}[leftmargin=*,itemsep=0pt,parsep=0pt,partopsep=0pt]
    \item In $A_{uu}$, each diagonal block $A_{uu}^{(k)}$ is a disjoint block--tridiagonal subsystem. 
    \item $ A_{ll}$ contains the diagonal blocks of the separators.
    \item Each $ A_{lu}^{(k)}$ is the coupling matrix corresponding to $ A_{uu}^{(k)}$, containing two non-zero blocks.
\end{itemize}

From (\ref{eq:reordered-eq1}), we can express the interior solution $ X_{u}$ in terms of the unknown separator solution $ X_{l}$: $ X_{u} =  A_{uu}^{-1}( B_{u} -  A_{lu}  X_{l})$. Substituting this into (\ref{eq:reordered-eq2}) eliminates $ X_{u}$ and yields a smaller system involving only the separators:
\begin{equation}
\label{eq:schur-system}
( A_{ll} -  A_{lu}^\top  A_{uu}^{-1}  A_{lu})  X_{l} =  B_{l} -  A_{lu}^\top  A_{uu}^{-1}  B_{u}
\end{equation}
The matrix $S =  A_{ll} -  A_{lu}^\top  A_{uu}^{-1}  A_{lu}$ is the Schur complement of $ A_{uu}$ in $\widetilde A$.
Since the Schur complement of an \gls*{spd} block-tridiagonal matrix is itself \gls*{spd} and block-tridiagonal, the reduced system can be recursively solved using the same procedure.


\subsection{Core Batched Kernels}
The recursive algorithm is built upon two core batched routines that operate on collections of independent block-tridiagonal systems: a batched factorization and a batched block triangular solve.

We use curly braces to denote a collection of independent problems that are solved in parallel. 
For instance, $\{A^{(k)}\}_{k=1}^K$ refers to a batch of $K$ block--tridiagonal matrices, 
and $\{B^{(k)}\}_{k=1}^K$ the corresponding right-hand sides. 
An assignment such as 
$\{C^{(k)}\} \gets \{f(A^{(k)},B^{(k)})\}$
indicates that the same operation $f$ is applied independently to each pair $(A^{(k)},B^{(k)})$. 
When the batch index $(k)$ is omitted, the notation should be read as applying to all members of the batch simultaneously.


\begin{algorithm}[h!]
  \caption{Batched Block-Cholesky Factorization}
  \label{alg:batch-factorize-sub}
  \begin{algorithmic}[1]
    \Require A batch of $K$ block-tridiagonal systems 
             $\{A^{(k)}\}_{k=1}^{K}$ each with $J$ diagonal blocks.
    \State \textit{-- Operations are executed in parallel across all $K$ systems --}
    \State $\{A_{1,1}^{(k)}\} \gets \operatorname{cholesky}(\{A_{1,1}^{(k)}\})$ \Comment{batched \texttt{potrf}}
    \For{$j = 1$ to $J-1$}
      \State $\{A_{j+1,j}^{(k)\top}\} \gets \{ (A_{j,j}^{(k)})^{-1} A_{j,j+1}^{(k)} \}$ \Comment{batched \texttt{trsm}}
      \State $\{A_{j+1,j+1}^{(k)}\} \gets \{ A_{j+1,j+1}^{(k)} - A_{j+1,j}^{(k)} A_{j+1,j}^{(k)\top} \}$ \\\Comment{batched \texttt{gemm}}
      \State $\{A_{j+1,j+1}^{(k)}\} \gets \{\operatorname{cholesky}(A_{j+1,j+1}^{(k)})\}$ \\ \Comment{batched \texttt{potrf}}
    \EndFor
  \end{algorithmic}
\end{algorithm}

\begin{algorithm}[h!]
  \caption{Batched Block-Cholesky Solve}
  \label{alg:batch-solve-sub}
  \begin{algorithmic}[1]
    \Require Factorized system matrix $\{A^{(k)}\}_{k=1}^{K}$; RHS 
    $\{B^{(k)}\}_{k=1}^{K}$
    \Ensure Solutions $\{X^{(k)}\}_{k=1}^{K}$
    \State \textit{-- Batched Forward Substitution --}
    \State $\{Y_1^{(k)}\} \gets \{ (A_{1,1}^{(k)})^{-1} B_1^{(k)} \}$ \Comment{batched \texttt{trsm}}
    \For{$j = 2$ to $J$}
      \State $\{\hat{B}_j^{(k)}\} \gets \{ B_j^{(k)} - A_{j,j-1}^{(k)} Y_{j-1}^{(k)} \}$ \Comment{batched \texttt{gemm}}
      \State $\{Y_j^{(k)}\} \gets \{ (A_{j,j}^{(k)})^{-1} \hat{B}_j^{(k)} \}$ \Comment{batched \texttt{trsm}}
    \EndFor
    \State \textit{-- Batched Backward Substitution --}
    \State $\{X_J^{(k)}\} \gets \{ (A_{J,J}^{(k)})^{-\top} Y_J^{(k)} \}$ \Comment{batched \texttt{trsm}}
    \For{$j = J-1$ to $1$}
      \State $\{\hat{Y}_j^{(k)}\} \gets \{ Y_j^{(k)} - A_{j+1,j}^{(k)\top} X_{j+1}^{(k)} \}$ \Comment{batched \texttt{gemm}}
      \State $\{X_j^{(k)}\} \gets \{ (A_{j,j}^{(k)})^{-\top} \hat{Y}_j^{(k)} \}$ \Comment{batched \texttt{trsm}}
    \EndFor
  \end{algorithmic}
\end{algorithm}

These batched kernels form the basic building blocks of our solver; in the following, we show how the recursive algorithm composes them into a hierarchy of factorization/solve steps.

\subsection{Recursive Factorization}
The factorization phase recursively computes all factors required for the solve. 
It computes and stores the factors for each independent interior block $ A_{uu}^{(k)}$ and for the Schur complement $S$ at each level of the hierarchy.
Each level involves three main computational stages:

\begin{enumerate}[leftmargin=*,itemsep=0pt,parsep=0pt,partopsep=0pt]
\item \textbf{Factorization of Interior Blocks.} 
We factorize $A_{uu}$ with the batched factorization routine (Alg. \ref{alg:batch-factorize-sub}). 

\item \textbf{Computation of intermediate factors.} 
To obtain the term $A_{uu}^{-1} A_{lu}$ in the Schur complement
$S = A_{ll} - A_{lu}^\top A_{uu}^{-1} A_{lu}$, 
we solve $A_{uu} F = A_{lu}$ using the batched solve routine (Alg. \ref{alg:batch-solve-sub}).

\item \textbf{Schur Complement Formation.} 
We form the update $A_{lu}^\top F$ and then call
\textsc{Assemble} to accumulate these segment-wise contributions into the global separator matrix $S$ (Alg. \ref{alg:schur-comp}).
\end{enumerate}

\begin{algorithm}[h!]
  \caption{Schur Complement Computation}
  \label{alg:schur-comp}
  \begin{algorithmic}[1]
    \Require Separator matrix $ A_{ll}$; Coupling matrix $ A_{lu}$; Intermediate Factor $F$.
    \Ensure Schur complement matrix $S$.
    \State $\{\hat{S}^{(k)}\} \gets \{ ( A_{lu}^{(k)})^\top F^{(k)} \}$ \Comment{batched \texttt{gemm}}
    \State $S \gets  A_{ll} - \operatorname{Assemble}(\{\hat S^{(k)}\})$
  \end{algorithmic}
\end{algorithm}

\begin{algorithm}[h!]
  \caption{Complete Parallel Recursive Factorization}
  \label{alg:recursive-factor}
  \begin{algorithmic}[1]
    \Require Block-tridiagonal matrix $A$.
    \Ensure Intermediate factor $F$; Factorized Schur Complement $S$.
    \If{$N \leq N_\star$} 
        \State \Call{SerialFactorize}{$A$} \Comment{Alg. \ref{alg:batch-factorize-sub} (K=1)}
        \State \Return
    \EndIf
    \State $ A_{uu},  A_{lu},  A_{ll} \gets$ \Call{Partition}{$A$}
    \State \Call{BatchedFactorize}{$ A_{uu}$} \Comment{Alg. \ref{alg:batch-factorize-sub}}
    \State $F \gets $\Call{BatchedSolve}{$ A_{uu},  A_{lu}$} \Comment{Alg. \ref{alg:batch-solve-sub}}
    \State $S \gets$ \Call{ComputeSchur}{$ A_{ll},  A_{lu}, F$} \Comment{Alg. \ref{alg:schur-comp}}
    \State \Call{RecursiveFactorize}{$S$} \Comment{Recursive call}
  \end{algorithmic}
\end{algorithm}

\subsubsection{The Complete Recursive Factorization.} The main recursive factorization algorithm in Alg. \ref{alg:recursive-factor} calls itself on the newly formed Schur complement until the base case is reached. We introduce a crossover parameter $N_\star$, defined as the maximum segment length for which a sequential block-Cholesky sweep is more efficient than introducing another level of recursion; whenever $N \leq N_\star$, the recursion terminates and the algorithm falls back to the serial factorization (i.e. Alg. \ref{alg:recursive-factor} with K = 1). 
The recursive factorization produces the hierarchy of factors needed to solve the system.


\subsection{Recursive Solve}
The solve phase unwinds the recursion, working from the separators on the coarsest level back to the interiors on the finest.
Each level also involves three main computational stages:




\begin{enumerate}[leftmargin=*,itemsep=0pt,parsep=0pt,partopsep=0pt]
\item \textbf{Separator right-hand side formation.}
The separator RHS is updated to
\(\hat{B}_l = B_l - A_{lu}^\top A_{uu}^{-1} B_u\),
with the intermediate factor \(F\) from the factorization phase (Alg. \ref{alg:schur-rhs}).

\item \textbf{Boundary Update for Interior Solves.}
Given \(X_l\), the interior right-hand sides are updated to
\(\hat{B}_u = B_u - A_{lu} X_l\) across all segments (Alg. \ref{alg:boundary-update}).

\item \textbf{Interior solves.}
The updated interior systems \(X_u\) are then solved using the
batched solve routine (Alg. \ref{alg:batch-solve-sub}).

\end{enumerate}

\begin{algorithm}[h!]
  \caption{Separator System RHS Computation}
  \label{alg:schur-rhs}
  \begin{algorithmic}[1]
    \Require Intermediate Factor $F$; 
    Interior RHS $B_u$; 
    Separator RHS $B_l$.
    \Ensure Updated separator RHS $\hat B_l$.
    \State $\{b^{(k)}\} \gets \{ - (F^{(k)})^\top B_u^{(k)} \}$ \Comment{batched \texttt{gemm}}
    \State $\hat B_l \gets B_l + \operatorname{Assemble}(\{b^{(k)}\})$
  \end{algorithmic}
\end{algorithm}


\begin{algorithm}[h!]
  \caption{Boundary Update}
  \label{alg:boundary-update}
  \begin{algorithmic}[1]
    \Require Interior RHS $B_u$; Separator solution $X_l$;  Coupling matrix $A_{lu}$.
    \Ensure Updated interior RHS $\hat B_u$.
    \State $\{\hat B_u^{(k)}\} \gets \{\, B_u^{(k)} - A_{lu}^{(k)} X_l^{(k)} \,\}$ \Comment{batched \texttt{gemm}}
  \end{algorithmic}
\end{algorithm}


\begin{algorithm}[h!]
  \caption{Complete Parallel Recursive Solve}
  \label{alg:recursive-solve}
  \begin{algorithmic}[1]
    \Require Factorized block-tridiagonal matrix $A$ at this recursion level 
             (with access to $A_{uu}, A_{lu}, A_{ll}$ and stored factors);
             Factored Schur complement $S$; Intermediate factor $F$; RHS $B$.
    \Ensure Solution $X$.
    \If{$N \le N_\star$}
        \State $X \gets \Call{SerialSolve}{A, B}$ \Comment{Alg. \ref{alg:batch-solve-sub} (K=1)}
        \State \textbf{return} $X$.
    \EndIf
    \State $ B_{u},  B_{l} \gets \Call{Partition}{B}$
    \State $\hat{B}_l \gets \Call{ComputeSeparatorRHS}{ F, B_{u},  B_{l}}$ \\ \Comment{Alg. \ref{alg:schur-rhs}}
    \State $ X_{l} \gets \Call{RecursiveSolve}{S, \hat{B}_l}$ \Comment{Recursive call}
    
    \State $\hat{B}_u \gets \Call{UpdateBoundary}{B_u,  X_{l}, F}$ \\\Comment{Alg. \ref{alg:boundary-update}}
    \State $ X_{u} \gets \Call{BatchedSolve}{A_{uu}, \hat{B}_u}$ \Comment{Alg. \ref{alg:batch-solve-sub}}
    \State $X \gets 
    \Call{Assemble}{ X_{u},  X_{l}}$
  \end{algorithmic}
\end{algorithm}

\subsubsection{The Complete Recursive Solve}
The main recursive routine in Alg. \ref{alg:recursive-solve} combines these steps into a full solve of the system.  
Starting from the original right-hand side \(B\), it recursively forms and solves separator systems and then recovers the interior variables \(X_u\) at each level.  
The \textsc{Assemble} step reverses the partitioning described in section \ref{sec:parallel-algo}, merging the interior and separator solutions into the final solution \(X\) in the original ordering.  
As in the factorization phase, the dominant work is expressed in terms of batched \gls*{blas}/\gls*{lapack} kernels, so each recursion level is executed through a small number of large, parallel GPU calls.

\section{Experiments}\label{sec:experiments}
\begin{table*}[h]
  \centering
  \caption{FP64 sweep for $Nn = 262144$, $n\in\{32,64,128,256,512,1024\}$. 
  Metrics: Factorization (Fact.) and solve runtime in milliseconds (ms). 
  Values are the mean of 10 runs.}
  \vspace{-2mm}
  \label{tab:fp64_sweep_clean}
  \small
  \setlength{\tabcolsep}{4pt}
  \begin{tabular}{|c c|c|ccccc|ccc|}
    \hline
    $N$ & $n$ & Phase & 
    \multicolumn{5}{c|}{\textbf{GPU method runtime (ms)}} & \multicolumn{3}{c|}{\textbf{CPU method runtime (ms)}} \\
    \cline{4-8}\cline{9-11}
     &  & & 
    \begin{tabular}[c]{@{}c@{}}cuDSS\\(NVIDIA)\end{tabular} &
    \begin{tabular}[c]{@{}c@{}}\tbdgpu\\(NVIDIA)\end{tabular} &
    \begin{tabular}[c]{@{}c@{}}Seq.\\(NVIDIA)\end{tabular} &
    \begin{tabular}[c]{@{}c@{}}\tbdgpu\\(AMD)\end{tabular} &
    \begin{tabular}[c]{@{}c@{}}Seq.\\(AMD)\end{tabular} &
    MA57 & CHOLMOD & LDL$^\top$ \\
    \hline

    \multirow{2}{*}{256} & \multirow{2}{*}{1024} 
      & Fact. & 298.02 & \textbf{273.35} & 391.91 & 356.50 & 841.33 & 104289.74 & 59434.58 & 247792.31 \\
      & & Solve       & 48.57  & \textbf{41.71}  & 224.02 & 113.55 & 212.11 & 658.92    & 428.91   & 1977.38   \\
    \hline

    \multirow{2}{*}{512} & \multirow{2}{*}{512} 
      & Fact. & 259.35 & \textbf{120.56} & 364.10 & 225.54 & 824.68 & 35962.12 & 25748.44 & 69248.17 \\
      & & Solve       & 55.36  & \textbf{41.53}  & 246.74 & 130.63 & 345.23 & 330.25   & 222.63   & 1023.62  \\
    \hline

    \multirow{2}{*}{1024} & \multirow{2}{*}{256} 
      & Fact. & 173.87 & \textbf{61.72}  & 405.98 & 224.81 & 873.97 & 13619.00 & 11085.06 & 15969.64 \\
      & & Solve       & \textbf{6.29}   & 21.70  & 352.26 & 24.23  & 430.72 & 172.14   & 116.98   & 515.54   \\
    \hline

    \multirow{2}{*}{2048} & \multirow{2}{*}{128} 
      & Fact. & \textbf{29.87}  & 41.91  & 517.58 & 245.73 & 1080.20& 5923.22  & 5127.79  & 4385.00  \\
      & & Solve       & \textbf{2.53}   & 30.86  & 566.19 & 42.36  & 721.62 & 94.93    & 72.29    & 285.93   \\
    \hline

    \multirow{2}{*}{4096} & \multirow{2}{*}{64} 
      & Fact. & \textbf{10.66}  & 12.65  & 737.65 & 11.93  & 1016.87& 2267.96  & 2031.70  & 1117.63  \\
      & & Solve       & \textbf{1.13}   & 6.84   & 953.29 & 12.00  & 1271.82& 42.78    & 34.71    & 111.78   \\
    \hline

    \multirow{2}{*}{8192} & \multirow{2}{*}{32} 
      & Fact. & \textbf{4.33}   & 7.95   & 1213.97& 6.49   & 1618.96& 1037.27  & 1069.48  & 357.45   \\
      & & Solve       & \textbf{0.68}   & 4.36   & 1779.63& 13.20  & 2549.24& 26.61    & 21.12    & 55.06    \\
    \hline

  \end{tabular}

\end{table*}

We evaluate \tbdgpu{} on large synthetic block--tridiagonal systems and a Kalman smoothing example, benchmarking against NVIDIA's sparse solver (cuDSS),  CPU solvers (CHOLMOD, MA57, LDLFactorizations.jl), and our sequential implementation (seq.), all in \gls*{fp64}. 
CPU experiments are run on an Intel Xeon Platinum 8562Y+ system (1007~GB RAM, Intel MKL), and GPU experiments on an NVIDIA H200 (141~GB HBM3e, CUDA~12.4) and an AMD MI300X (128~GB HBM3e, ROCm~6.4.1).
$N_*$ is determined dynamically for each recursive level in each run.
We report factorization and solve times and the residual norm $\|Ax-b\|_2$; all numbers are averaged over 10 runs. The code is provided in the previous footnote.

\subsection{Synthetic Experiment}

We use randomly generated, well-conditioned \gls*{spd} block--tridiagonal matrices so that conditioning effects are separated from computational performance. 
For a fixed total dimension $Nn = 262{,}144$, we sweep over different $(N,n)$ pairs as summarized in Table \ref{tab:fp64_sweep_clean}. 
All solvers achieve residual norms at least on the order of $10^{-9}$.


\subsubsection{GPU vs.\ CPU baselines}  
Across all problem sizes, GPU-based solvers outperform CPU solvers by one to three orders of magnitude in time-to-solution. In our experiment, CPU methods are competitive only for small blocks ($n \leq 64$), while for larger blocks the GPU runtimes remain in the sub-second regime.

\subsubsection{\tbdgpu{} vs.\ cuDSS}  
For small to moderate block sizes ($n \leq 128$), cuDSS attains very low runtimes by exploiting highly optimized sparse kernels. 
As $n$ grows, \tbdgpu{} becomes increasingly favorable: for $(N,n)=(1024,256)$, total time to solution is $83$~ms versus $180$~ms for cuDSS, and at $(4096,64)$ it stays within a factor of two of cuDSS. 
The advantage of \tbdgpu{} is most pronounced for large blocks ($n \geq 512$), where its batched dense-kernel formulation can fully utilize memory bandwidth and tensor cores.

\subsection{Kalman Smoothing Experiment}
Kalman smoothing arises in applications like navigation, tracking, and signal processing when estimating the trajectory of a hidden state from noisy measurements. 
The system evolves according to
\begin{align}
x_k &= G_k x_{k-1} + w_k, &k=1,\dots,N, \\
z_k &= H_k x_k + v_k, &k=1,\dots,N ,
\end{align}
where $x_k \in \mathbb{R}^n$ is the hidden state, 
$z_k \in \mathbb{R}^m$ is the noisy measurement, 
$G_k \in \mathbb{R}^{n \times n}$ is the state transition matrix,
and $H_k \in \mathbb{R}^{m \times n}$ is the observation matrix at time $k$.
The process and measurement noises, 
$w_k \sim \mathcal{N}(0,Q_k)$ and $v_k \sim \mathcal{N}(0,R_k)$, 
are independent with known covariances. 
The initial condition $x_0$ is assumed known, with $G_1 = I_n$.

Estimating the most likely sequence of states 
given all measurements is known as the
\gls*{map} smoothing problem.  
Stacking all states $x^T = [x_1^T,\dots,x_N^T]$ and all measurements
$z^T = [z_1^T,\dots,z_N^T]$, this is equivalent to minimizing the negative log-posterior,

\[
\min_x\;\tfrac{1}{2}\|Hx-z\|_{R^{-1}}^2
+ \tfrac{1}{2}\|Gx-\zeta\|_{Q^{-1}}^2 ,
\]
where the matrices $H$, $R$, $Q$, $G$, and the vector $\zeta$ 
are defined as in \cite{ARAVKIN2021821}. 
The corresponding optimality conditions yield an
\gls*{spd} block-tridiagonal system:
\begin{equation}
\label{eq:ksmooth-btd}
(H^\top R^{-1}H + G^\top Q^{-1}G)\,x =
H^\top R^{-1}z + G^\top Q^{-1}\zeta .
\end{equation}

\noindent \noindent The system can be written compactly as
$A x= b,$ where the coefficient matrix in (\ref{eq:ksmooth-btd}) has $n \times n$ blocks
\begin{align*}
A_{k,k}   &= Q_k^{-1} + G_{k+1}^\top Q_{k+1}^{-1} G_{k+1}
             + H_k^\top R_k^{-1} H_k, \\
A_{k+1,k} &= -\,Q_{k+1}^{-1} G_{k+1},
\end{align*}
with the convention $G_{N+1}=0$, and the right-hand side blocks are
$
b_k = H_k^\top R_k^{-1} z_k + G_k^\top Q_k^{-1} \zeta_k .
$ 
This system has exactly the block-tridiagonal \gls*{spd} structure. Classical \gls*{rts} and related smoothers can be viewed as particular block-elimination schemes for (\ref{eq:ksmooth-btd}); here we instead solve the normal equations directly with our solver.

\subsubsection{Neuroscience experiment}
To illustrate performance in a realistic high-dimensional regime, we draw inspiration from neural population dynamics in motor cortex, where latent activity exhibits smooth rotational structure and is observed through a large number of noisy spike trains~\citep{sabatini2024reach, churchland2012neural}.  
We instantiate the model with latent state dimension $n=256$, observation dimension $m=1024$, and horizon $N=100$ (10\,s sampled at $\Delta t = 0.1$\,s). 
The latent dynamics are modeled by constructing $G$ from damped two-dimensional rotations. The observation matrix $H$ is tall and well-conditioned. The process noise covariance $Q$ is dense and symmetric positive definite, while the observation noise covariance $R$ is diagonal and scaled larger than $Q$ to reflect more reliable latent dynamics than sensor observations.


\subsubsection{Evaluation}
We report (i) the residual norm $\|Ax-b\|_2$ and (ii) solver wall-clock times for initialization, factorization, and solution phases.  
Both \tbdgpu{} and cuDSS converge to the same residual level of $\sim 6\times 10^{-12}$, confirming numerical equivalence.  
The measured times (in milliseconds) are as follows:
\begin{center}
\begin{tabular}{lcccc}
\toprule
Solver & Init & Factor & Solve & Total \\
\midrule
\tbdgpu{} & 109.6 & 13.3 & 8.6 & 21.9\\
cuDSS   & 246.5 & 27.5 & 1.8 & 29.3 \\
\bottomrule
\end{tabular}
\end{center}



On this tall-observation, large-state problem, \tbdgpu{} avoids general sparse overhead and achieves a clear runtime advantage over cuDSS. Although this example is modest in size, our synthetic benchmarks indicate that the performance gap in favor of \tbdgpu{} widens further as the block size and horizon increase. 

\section{Conclusions}\label{sec:discussion}

We have presented the \tbdgpu{} solver, a GPU implementation for large-scale block-tridiagonal linear systems. By combining a recursive Schur complement strategy with batched dense linear algebra kernels, our approach overcomes the inherent sequential bottlenecks of traditional algorithms while maximizing GPU utilization.
On synthetic block-tridiagonal problems, \tbdgpu{} consistently outperforms sequential GPU solvers while remaining competitive compared to NVIDIA's cuDSS on large block sizes.
The structure-aware design of \tbdgpu{} allows it to exploit BLAS and LAPACK batched kernels more effectively than general-purpose sparse solvers.
However, since \tbdgpu{} relies on sequential calls to batched routines at each recursion level, it requires sufficiently large block sizes to amortize kernel launch overheads.
In the future, we plan to develop a solver that fully fuses the recursive algorithm into a single GPU kernel, to minimize the kernel launch overhead and further improve performance on smaller block sizes.

\section*{DECLARATION OF GENERATIVE AI AND AI-ASSISTED TECHNOLOGIES IN THE WRITING PROCESS}
During the preparation of this work the author(s) used ChatGPT in order to correct grammar and polish writing. After using this tool/service, the author(s) reviewed and edited the content as needed and take(s) full responsibility for the content of the publication.

\bibliography{references}

@InProceedings{dongarra2017bblas,
author="Dongarra, Jack
and Hammarling, Sven
and Higham, Nicholas J.
and Relton, Samuel D.
and Zounon, Mawussi",
title="Optimized Batched Linear Algebra for Modern Architectures",
booktitle="Euro-Par 2017: Parallel Processing",
year="2017",
publisher="Springer International Publishing",
address="Cham",
pages="511--522",
isbn="978-3-319-64203-1"
}

@Inbook{Aravkin2014,
author="Aravkin, Aleksandr Y.
and Burke, James V.
and Pillonetto, Gianluigi",
title="Optimization Viewpoint on Kalman Smoothing with Applications to Robust and Sparse Estimation",
bookTitle="Compressed Sensing {\&} Sparse Filtering",
year="2014",
publisher="Springer Berlin Heidelberg",
address="Berlin, Heidelberg",
pages="237--280",
isbn="978-3-642-38398-4",
}

@article{ARAVKIN2021821,
title = {Algorithms for Block Tridiagonal Systems: Stability Results for Generalized Kalman Smoothing},
journal = {IFAC-PapersOnLine},
volume = {54},
number = {7},
pages = {821-826},
year = {2021},
note = {19th IFAC Symposium on System Identification SYSID 2021},
issn = {2405-8963},
author = {Aleksandr Y. Aravkin and James V. Burke and Bradley M. Bell and Gianluigi Pillonetto}
}

@inproceedings{GanderGolub1997CyclicReduction,
  author    = {Gander, Walter and Golub, Gene H.},
  title     = {Cyclic Reduction -- History and Applications},
  booktitle = {Proceedings of the Workshop on Scientific Computing},
  location  = {Hong Kong},
  publisher = {Springer-Verlag},
  year      = {1997},
  url       = {https://people.inf.ethz.ch/gander/papers/cyclic.pdf},
}

@article{hockney1965cr,
author = {Hockney, R. W.},
title = {A Fast Direct Solution of Poisson's Equation Using Fourier Analysis},
year = {1965},
issue_date = {Jan. 1965},
publisher = {Association for Computing Machinery},
address = {New York, NY, USA},
volume = {12},
number = {1},
issn = {0004-5411},
journal = {J. ACM},
month = jan,
pages = {95–113},
numpages = {19}
}

@article{magma2024,
author = {Ahmad Abdelfattah and Natalie Beams and Robert Carson and Pieter Ghysels and Tzanio Kolev and Thomas Stitt and Arturo Vargas and Stanimire Tomov and Jack Dongarra},
title ={MAGMA: Enabling exascale performance with accelerated BLAS and LAPACK for diverse GPU architectures},

journal = {The International Journal of High Performance Computing Applications},
volume = {38},
number = {5},
pages = {468-490},
year = {2024}
}

@misc{NVIDIACUBLAS2025,
  author       = {{NVIDIA Corporation}},
  title        = {cuBLAS Library User Guide},
  year         = {2025},
  howpublished = {\url{https://docs.nvidia.com/cuda/cublas/}}
}

@misc{NVIDIACUDSS2025,
  author       = {{NVIDIA Corporation}},
  title        = {cuDSS: Direct Sparse Solvers},
  year         = {2025},
  howpublished = {\url{https://docs.nvidia.com/cuda/cudss/}}
}

@misc{NVIDIACUSOLVER2025,
  author       = {{NVIDIA Corporation}},
  title        = {cuSOLVER Library},
  year         = {2025},
  howpublished = {\url{https://docs.nvidia.com/cuda/cusolver/}}
}

@misc{NVIDIACUSPARSE2025,
  author       = {{NVIDIA Corporation}},
  title        = {cuSPARSE Library User Guide},
  year         = {2025},
  howpublished = {\url{https://docs.nvidia.com/cuda/cusparse/}}
}

@article{kkpc2020,
    title = "PaScaL\_TDMA: A library of parallel and scalable solvers for massive tridiagonal system",
    author = "Kim, Ki-Ha and Kang, Ji-Hoon and Pan, Xiaomin and Choi, Jung-Il",
    journal = "Computer Physics Communications",
    volume = "260",
    pages = "107722",
    year = "2021",
    issn = "0010-4655",
}

@article{tolmachev2025pfsolve,
author = {Tolmachev, Dmitrii and Marti, Philippe and Castiglioni, Giacomo and Jackson, Andrew},
title = {High Performance Solution of Tridiagonal Systems on the GPU},
year = {2025},
issue_date = {June 2025},
publisher = {Association for Computing Machinery},
address = {New York, NY, USA},
volume = {12},
number = {2},
issn = {2329-4949},
journal = {ACM Trans. Parallel Comput.},
month = may,
articleno = {5},
numpages = {25}
}

@misc{du2025gatogpuacceleratedbatchedtrajectory,
      title={GATO: GPU-Accelerated and Batched Trajectory Optimization for Scalable Edge Model Predictive Control}, 
      author={Alexander Du and Emre Adabag and Gabriel Bravo and Brian Plancher},
      year={2025},
      eprint={2510.07625},
      archivePrefix={arXiv},
      primaryClass={cs.RO},
      url={https://arxiv.org/abs/2510.07625}, 
}

@techreport{Thomas1949,
  author      = {L. H. Thomas},
  title       = {Elliptic Problems in Linear Differential Equations over a Network},
  institution = {Watson Scientific Computing Laboratory, Columbia University},
  year        = {1949},
  note        = {Unpublished Report},
  url         = {https://www.scirp.org/reference/referencespapers?referenceid=1807055}
}

@article{Rauch1965,
  author  = {H. E. Rauch and F. Tung and C. T. Striebel},
  title   = {Maximum Likelihood Estimates of Linear Dynamic Systems},
  journal = {AIAA Journal},
  year    = {1965},
  volume  = {3},
  number  = {8},
  pages   = {1445--1450},
}

@misc{ROCm2025,
  author       = {{Advanced Micro Devices, Inc.}},
  title        = {{ROCm} Documentation},
  year         = {2025},
  howpublished = {\url{https://rocm.docs.amd.com/}}
}

@article{klein2023tridigpu,
author = {Klein, Christoph and Strzodka, Robert},
title = {Tridigpu: A GPU Library for Block Tridiagonal and Banded Linear Equation Systems},
year = {2023},
issue_date = {March 2023},
publisher = {Association for Computing Machinery},
address = {New York, NY, USA},
volume = {10},
number = {1},
issn = {2329-4949},
journal = {ACM Trans. Parallel Comput.},
month = mar,
articleno = {4},
numpages = {33}
}

@inproceedings{pacaudGPUacceleratedDynamicNonlinear2024,
  title = {{{GPU-accelerated}} Dynamic Nonlinear Optimization with {{ExaModels}} and {{MadNLP}}},
  booktitle = {2024 {{IEEE}} 63rd {{Conference}} on {{Decision}} and {{Control}} ({{CDC}})},
  author = {Pacaud, Fran{\c c}ois and Shin, Sungho},
  year = {2024},
  month = {dec},
  pages = {5963--5968},
  issn = {2576-2370},
  urldate = {2025-08-05},
}

@inproceedings{rennichAcceleratingSparseCholesky2014,
  title = {Accelerating {{Sparse Cholesky Factorization}} on {{GPUs}}},
  booktitle = {2014 4th {{Workshop}} on {{Irregular Applications}}: {{Architectures}} and {{Algorithms}} ({{IA}}{\textasciicircum}3)},
  author = {Rennich, Steven C. and Stosic, Darko and Davis, Timothy A.},
  year = {2014},
  month = {nov},
  pages = {9--16},
  issn = {2767-942X},
  urldate = {2025-08-26},
}

@article{wrightPartitionedDynamicProgramming1991,
  title = {Partitioned {{Dynamic Programming}} for {{Optimal Control}}},
  author = {Wright, Stephen J.},
  year = {1991},
  month = {nov},
  journal = {SIAM Journal on Optimization},
  volume = {1},
  number = {4},
  pages = {620--642},
  publisher = {{Society for Industrial and Applied Mathematics}},
  issn = {1052-6234},
  urldate = {2024-01-17},
}

@book{Smith1996,
author = {Smith, Barry F. and Bj\o{}rstad, Petter E. and Gropp, William D.},
title = {Domain decomposition: parallel multilevel methods for elliptic partial differential equations},
year = {1996},
isbn = {052149589X},
publisher = {Cambridge University Press},
address = {USA}
}

@article{TEREKHOV2013245,
title = {A fast parallel algorithm for solving block-tridiagonal systems of linear equations including the domain decomposition method},
journal = {Parallel Computing},
volume = {39},
number = {6},
pages = {245-258},
year = {2013},
issn = {0167-8191},
author = {Andrew V. Terekhov}
}

@article{churchland2012neural,
  author    = {Churchland, Mark M. and Cunningham, John P. and Kaufman, Matthew T. and Foster, Justin D. and Nuyujukian, Paul and Ryu, Stephen I. and Shenoy, Krishna V.},
  title     = {Neural population dynamics during reaching},
  journal   = {Nature},
  year      = {2012},
  volume    = {487},
  number    = {7405},
  pages     = {51--56},
}

@article{sabatini2024reach,
  author    = {Sabatini, David A. and Kaufman, Matthew T.},
  title     = {Reach-dependent reorientation of rotational dynamics in motor cortex},
  journal   = {Nature Communications},
  year      = {2024},
  volume    = {15},
  number    = {1},
  pages     = {7007},
}

@article{belov2017,
year = {2017},
month = {nov},
publisher = {IOP Publishing},
volume = {929},
number = {1},
pages = {012035},
author = {Belov, P A and Nugumanov, E R and Yakovlev, S L},
title = {The arrowhead decomposition method for a block-tridiagonal system of linear equations},
journal = {Journal of Physics: Conference Series}
}

@article{sweet1977,
author = {Sweet, Roland A.},
title = {A Cyclic Reduction Algorithm for Solving Block Tridiagonal Systems of Arbitrary Dimension},
journal = {SIAM Journal on Numerical Analysis},
volume = {14},
number = {4},
pages = {706-720},
year = {1977},
}

\end{document}